\documentclass[12pt]{article}

\usepackage{graphicx}
\begin{document}

\begin{center}
{\bf Dyonic and magnetic black holes with nonlinear arcsin-electrodynamics } \\
\vspace{5mm} S. I. Kruglov
\footnote{E-mail: serguei.krouglov@utoronto.ca}
\underline{}
\vspace{3mm}

\textit{ Department of Physics, University of Toronto, \\60 St. Georges St.,
Toronto, ON M5S 1A7, Canada\\
Department of Chemical and Physical Sciences,\\ University of Toronto Mississauga,\\
3359 Mississauga Rd. N., Mississauga, ON L5L 1C6, Canada}
\vspace{5mm}
\end{center}
\begin{abstract}
Dyonic and magnetic black holes solutions with spherically symmetric configurations in general relativity are  obtained. Black holes possessing electric and magnetic charges are studied where the source of the gravitational field is electromagnetic fields obeying the arcsin-electrodynamics. We find corrections to Coulomb's law and Reissner$-$ Nordstr\"{o}m solutions. The principles of causality and unitarity are investigated. We obtain the Hawking temperature and it was shown that at some event horizons there are second-order phase transitions.
\end{abstract}

\section{Introduction}

After the Hawking works \cite{Hawking}, \cite{Hawking1}, physics of black holes (BHs) and their thermodynamics attract much attention. Dyonic BHs (with both magnetic and electric charges) represent the special more complicated class of BHs as compared with pure magnetic and pure electric BHs.
Dyonic solutions in black holes were found in the string theory \cite{Mignemi}, \cite{Mignemi0}, \cite{Jatkar}, supergravity \cite{Chamseddine}, \cite{Chow}, gravity's rainbow \cite{Panahiyan}, massive gravity \cite{Hendi} and in many other models. It is worth noting that dyonic BHs have applications in different areas such as condensed matter physics and thermodynamics. Thus, the Hall conductivity in the framework of AdS/CFT correspondence was investigated in \cite{Hartnoll}. The results found are in agreement with the predictions of the hydrodynamic analysis. The Nernst effect in the framework of the dyonic BH was described in \cite{Hartnoll1}. Making use of the AdS/CFT correspondence, it was shown that dyonic BHs in AdS spacetime are dual to stationary solutions of the equations of relativistic magnetohydrodynamics \cite{Caldarelli}. The superconductivity was studied in \cite{Albash} and thermodynamic properties of the dyonic BH were investigated in \cite{Dutta}. All these show the importance of studying dyonic BHs.

In this paper we investigate the dyonic BH in the framework of the nonlinear arcsin-electrodynamics.
The nonlinear electrodynamics (NED) can solve the problems of singularities in the origin of particles and the problem of infinite self-energy at the classical level.  Born and Infeld (BI) \cite{Born} proposed the model of NED that can solve problems of singularities. Heisenberg and Euler have shown that QED due to loop corrections gives NED \cite{Heisenberg}. Then some NED models were appeared that possess similar properties \cite{Soleng}-\cite{Kruglov1}. NED's coupled to general relativity (GR) were studied in \cite{Pellicer}-\cite{Quiros}. Thermodynamics of the BH and corrections to Reisner$-$Nordstr\"{o}m (RN) solutions   were investigated \cite{Hendi1}-\cite{Kruglov2}. Electrically and magnetically charged BHs were studied \cite{Yajima}-\cite{Kruglov3}. It was demonstrated that phase transitions can occur in the BH.
Inflation and current acceleration of the universe also can be explained by  NED coupled to GR  \cite{Garcia}-\cite{Kruglov4}.

The paper is organized as follows. The principles of causality and unitarity of arcsin-electrodynamics are considered in section 2. In section 3 the dyonic solution of the BH is obtained. Corrections to Coulomb's law and RN solutions are found. The curvature singularities are studied and we calculate the Kretschmann scalar. The thermodynamics of the BH is considered in section 4. The Hawking temperature of the BH is obtained. In section 5 we study the magnetically charged BH. We show that at some model parameters there can be naked singularities, extremal BH solutions and BH solutions with two horizons.  The Hawking temperature and corrections to RN solution are found. We show the possibility of phase transitions in the BH. Section 6 is devoted to a conclusion.

We use units with $c=1$, $k_B=1$, and signature $\eta=\mbox{diag}(-,+,+,+)$.

\section{The model and principles of causality and unitarity}

The Lagrangian density of arcsin-electrodynamics was proposed in \cite{Krug}, \cite{Krug6} (see also \cite{Krug1}) and is given by
\begin{equation}
{\cal L} = -\frac{1}{\beta}\arcsin(\beta{\cal F}),
 \label{1}
\end{equation}
where ${\cal F}=(1/4)F^{\mu\nu}F_{\mu\nu}=(\textbf{B}^2-\textbf{E}^2)/2$ and $F_{\mu\nu}=\partial_\mu A_\nu-\partial_\nu A_\mu$ is the field strength tensor. The parameter $\beta$ possesses the dimension of (length)$^{4}$ and $\beta\cal F$ is dimensionless. At ${\cal F}\rightarrow 0$ the Lagrangian density (1) becomes the Maxwell Lagrangian density ($-{\cal F}$).
In this section we consider arcsin-electrodynamics in the framework of special relativity.
The maximum of the electric field at the origin is finite and equals $E_{max}=\sqrt{2/\beta}$ as well as in BI electrodynamics. The self-energy of point-like charges is also finite \cite{Krug1}.

 Principles of causality and unitarity hold when inequalities are satisfied \cite{Shabad2},
\begin{equation}
 {\cal L}_{\cal F}\leq 0,~~~~{\cal L}_{{\cal F}{\cal F}}\geq 0,~~~~
{\cal L}_{\cal F}+2{\cal F} {\cal L}_{{\cal F}{\cal F}}\leq 0,
\label{2}
\end{equation}
where ${\cal L}_{\cal F}\equiv\partial{\cal L}/\partial{\cal F}$. Making use of Eq. (1) we obtain
\[
{\cal L}_{\cal F}= -\frac{1}{\sqrt{1-(\beta{\cal F})^2}},~~~~ {\cal L}_{{\cal F}{\cal F}}=-\frac{\beta^2{\cal F}}{[1-(\beta{\cal F})^2]^{3/2}},
\]
\begin{equation}
{\cal L}_{\cal F}+2{\cal F} {\cal L}_{{\cal F}{\cal F}}=-\frac{1+(\beta{\cal F})^2}{[1-(\beta{\cal F})^2]^{3/2}}.
\label{3}
\end{equation}
We assume that $\beta\geq 0$ and $\beta{\cal F}<1$.
One finds from Eqs. (2) and (3) that the principle of causality (${\cal L}_{{\cal F}{\cal F}}\geq 0$) takes place if $|\textbf{E}|\geq |\textbf{B}|$.
It should be noted that in BI electrodynamics for any electric and magnetics fields the principles of causality and unitarity hold.
In the following we consider fields described by NED as classical. Gravitational fields are also treated as classical as quantum gravity is not developed yet.
It was shown in \cite{Kruglov5} that in the model of holographic s-wave superconductors
with arcsin-electrodynamics the condensation formation depends on the parameter $\beta$ weakly as compared with the model with BI electrodynamics. In addition, to create the condensation is easier in arcsin-electrodynamics comparing to BI electrodynamics. One can consider the arcsin-electrodynamics as a toy-model that allows us to obtain dyonic solutions.

The symmetric energy-momentum tensor corresponding to arcsin-electrodynamics is given by
\begin{equation}
T_{\mu}^{\nu}=-\frac{F_\mu^\alpha F_\alpha^\nu}{\sqrt{1-(\beta {\cal F})^2}} -\delta^\nu_\mu{\cal L}.
\label{4}
\end{equation}
From Eq. (4) we obtain the energy density
\begin{equation}
\rho=T_{0}^{0}=\frac{E^2}{\sqrt{1-(\beta {\cal F})^2}} +\frac{1}{\beta}\arcsin(\beta{\cal F}).
\label{5}
\end{equation}

\section{Dyonic solution}

We start with arcsin-electrodynamics coupled to GR with the action
\begin{equation}
I=\int d^4x\sqrt{-g}\left(\frac{1}{16\pi G}R+ {\cal L}\right),
\label{6}
\end{equation}
where $G$ is Newton's constant and ${\cal L}$ is given by Eq. (1). We assume that the metric is static, spherically symmetric and is given by the line element:
\begin{equation}
ds^2=-A(r)dt^2+\frac{1}{A(r)}dr^2+r^2(d\vartheta^2+\sin^2\vartheta d\phi^2).
\label{7}
\end{equation}
Varying action (6) with respect to the metric tensor $g_{\mu\nu}$ and the field tensor $F_{\mu\nu}$ we obtain equations as follows:
\begin{equation}
R_{\mu\nu}-\frac{1}{2}g_{\mu\nu}R=8\pi G\left({\cal L}_{\cal F}F_\mu^{~\alpha}F_{\nu\alpha}-g_{\mu\nu}{\cal L}\right),
\label{8}
\end{equation}
\begin{equation}
\partial_\mu\left(\sqrt{-g}F^{\mu\nu}{\cal L}_{\cal F}\right)=0.
\label{9}
\end{equation}
From Eq. (9), making use of the radial electric field $E=F_{0r}$, we obtain
\begin{equation}
\partial_r\left(r^2E{\cal L}_{\cal F}\right)=0,
\label{10}
\end{equation}
with the solution $r^2E{\cal L}_{\cal F}=q_e$ where $q_e$ is the integration constant which we identified with the electric charge. By using the Bianchi identities $\nabla_\mu \tilde{F}_{\mu\nu}=0$, where $\tilde{F}_{\mu\nu}$ is a dual tensor, and $B=F_{\theta\phi}$ is a radial magnetic induction field, one finds
\begin{equation}
\partial_r(r^2 B)=0.
\label{11}
\end{equation}
The solution to Eq. (11) is $r^2B=q_m$ (see also \cite{Bronnikov3}, \cite{Bronnikov4}\footnote{We use other  notations comparing to \cite{Bronnikov3}, \cite{Bronnikov4}.}), where the constant of integration is the magnetic charge $q_m$.
As a result, field equations give
\begin{equation}
B^2=\frac{q^2_m}{r^4},~~~~E^2=\frac{q_e^2}{{\cal L}^2_{\cal F}r^4}=\frac{q^2_e}{r^4}\left(1-(\beta{\cal F})^2\right),
\label{12}
\end{equation}
\begin{equation}
{\cal F}=\frac{q_m^2}{2r^4}-\frac{q_e^2}{2r^4}\left(1-(\beta{\cal F})^2\right).
\label{13}
\end{equation}
The solution to quadratic equation (13) for $|E|>|B|$ is
\begin{equation}
{\cal F}=\frac{1}{\beta^2q_e^2}\left(r^4-\sqrt{r^8+\beta^2q_e^2(q_e^2-q_m^2)}\right).
\label{14}
\end{equation}
For this case ($|E|>|B|$) the principles of causality and unitarity hold. From Eq. (14) we obtain the dyonic solution
\begin{equation}
E^2=\frac{2}{\beta^2q_e^2}\left(\sqrt{r^8+\beta^2q_e^2(q_e^2-q_m^2)}-r^4\right)+\frac{q_m^2}{r^4}.
\label{15}
\end{equation}
For the case $q_e=q_m$, one finds from Eq. (15) $E=q_e/r^2$. When $q_m\neq 0$ the electric field at the origin ($r=0$) has a singularity. However, if the magnetic charge is zero ($q_m=0$), the singularity is absent.
Making use of Eq. (15) we obtain at $r\rightarrow\infty$
\begin{equation}
E=\frac{q_e}{r^2}-\frac{\beta^2q_e(q_m^2-q_e^2)^2}{8r^{10}}+{\cal O}(r^{-15}).
\label{16}
\end{equation}
It follows from Eq. (16) that corrections to Coulomb's law are in the order of ${\cal O}(r^{-10})$. For the self-dual solution, $q_m=q_e$, corrections to Coulomb's law vanish.

Let us compare corrections to Coulomb's law in our model with corrections in BI electrodynamics. The Lagrangian density in BI electrodynamics is given by
\begin{equation}
{\cal L} =\frac{1}{\beta}\left(1-\sqrt{1 +2\beta{\cal F}}\right),
 \label{17}
\end{equation}
where we have omitted the term $G=F_{\mu\nu}\tilde{F}^{\mu\nu}/4=\textbf{E}\cdot \textbf{B}$. Such term $G$ can be included in arcsin-electrodynamics \cite{Krug1}. From Eq. (17) we obtain
\[
{\cal L}_{\cal F}=-\frac{1}{\sqrt{1+2\beta{\cal F}}},~~~B^2=\frac{q^2_m}{r^4},~~~~E^2=\frac{q_e^2}{{\cal L}^2_{\cal F}r^4}=\frac{q^2_e}{r^4}\left(1+2\beta{\cal F}\right),
\]
\begin{equation}
{\cal F}=\frac{q_m^2-q_e^2}{2(r^4+\beta q_e^2)}.
\label{18}
\end{equation}
Making use of Eq. (18) we find the electric field
\begin{equation}
E=\frac{q_e}{r^2}\sqrt{\frac{r^4+\beta q_m^2}{r^4+\beta q_e^2}}.
\label{19}
\end{equation}
It follows from Eq. (19) that at the self-dual case, $q_e=q_m$, corrections to Coulomb's law are absent and $E=q_e/r^2$. Thus, the same feature takes place in arcsin-electrodynamics. From Eq. (19) we obtain the asymptotic value at $r\rightarrow\infty$
\begin{equation}
E=\frac{q_e}{r^2}-\frac{\beta q_e(q_e^2-q_m^2)}{2r^{6}}
+\frac{\beta^2q_e(q_e^2-q_m^2)(3q_e^2+q_m^2)}{8r^{10}}+{\cal O}(r^{-14}).
\label{20}
\end{equation}
Equation (20) shows that corrections to Coulomb's law in BI electrodynamics are in the order of ${\cal O}(r^{-6})$. But in arsin-electrodynamics corrections to Coulomb's law are ${\cal O}(r^{-10})$ and, therefore, arsin-electrodynamics is ``closer" to Maxwell's electrodynamics at large $r$. It is worth noting that the effect of birefringence is absent in the model under consideration \cite{Krug1} as well as in BI electrodynamics. These show that arcsin-electrodynamics is of theoretical interest.

According to the Einstein equations the metric function $A(r)$ in Eq. (7) reads \cite{Bronnikov}
\begin{equation}
A(r) = 1-\frac{2M(r)G}{r}.
\label{21}
\end{equation}
The mass function $M(r)$ is given by
\begin{equation}
M(r)=m-\int_r^\infty\rho(r)r^2dr,
\label{22}
\end{equation}
where $m$ is the total mass (including an electromagnetic mass) which represents the free parameter. From Eqs. (5) and (14) we obtain the energy density for dyonic configuration
\[
\rho(r)=\frac{1}{\beta r^4}\sqrt{\beta^2q_e^2q_m^2+2r^4(\sqrt{r^8+\beta^2q_e^2(q_e^2-q_m^2)}-r^4)}
\]
\begin{equation}
-\frac{1}{\beta}\arcsin\left(\frac{\sqrt{r^8+\beta^2q_e^2(q_e^2-q_m^2)}-r^4}{\beta q_e^2}\right).
\label{23}
\end{equation}
For convenience we introduce the dimensionless variables $z=r^4/(\beta q_e^2)$, $n=q_m^2/q_e^2$. Then Eq. (23) becomes
\begin{equation}
\beta\rho(z)=\frac{1}{z}\sqrt{n+2z(\sqrt{z^2+1-n}-z)}
-\arcsin\left(\sqrt{z^2+1-n}-z\right).
\label{24}
\end{equation}
With the help of Eqs. (22) and (24) we obtain
\[
M(z)=m-\frac{q_e^{3/2}}{4\beta^{1/4}}\int_z^\infty\biggl(\frac{1}{z}\sqrt{n+2z(\sqrt{z^2+1-n}-z)}
\]
\begin{equation}
-\arcsin\left(\sqrt{z^2+1-n}-z\right)\biggr)\frac{dz}{z^{1/4}}.
\label{25}
\end{equation}
It should be noted that expression (25) for the mass function is valid for any $z\neq 0$ (or $r\neq 0$) and it is finite. If $z=0$ (or $r=0$) in Eqs. (22) and (25) then $M(0)=m_{el}=\infty$ ($m_{el}$ is an electromagnetic mass).
Thus, the singularity at $r=0$ is still present. To evaluate the mass function at $z\rightarrow\infty$ ($r\rightarrow\infty$) we explore the $\rho$ in Eq. (24) at $z\rightarrow\infty$,
\begin{equation}
\beta\rho(z)=\frac{1+n}{2z}+{\cal O}(z^{-3}).
\label{26}
\end{equation}
Then from Eq. (25) one obtains the mass function at $r\rightarrow\infty$,
\begin{equation}
M(r)=m-\frac{q_e^2+q_m^2}{2r}+{\cal O}(r^{-9}).
\label{27}
\end{equation}
Making use of Eqs. (21) and (27) we find an asymptotic of the metric function at $r\rightarrow\infty$,
\begin{equation}
A(r)=1-\frac{2mG}{r}+\frac{(q_e^2+q_m^2)G}{r^2}+{\cal O}(r^{-10}).
\label{28}
\end{equation}
 Equation (28) shows that corrections to the Reissner$-$Nordstr\"{o}m solution in our model are in the order of ${\cal O}(r^{-10})$. It follows from Eq. (24) that at $q_e=q_m$ ($n=1$) we have exactly $\beta\rho(z)=1/z$ and corrections to the Reissner$-$Nordstr\"{o}m solution are absent. The analysis of the metric function can be performed with the help of the exact mass function (25), which is valid for any $r\neq 0$, and the definition of $A(r)$ given in Eq. (21). We will make this analysis in Sec. 5 for the magnetic BH.

Let us investigate the possible BH curvature singularities. For this purpose we consider the Kretschmann scalar $K(r)$ which is given by \cite{Hendi1}
\begin{equation}\label{29}
 K(r)\equiv R_{\mu\nu\alpha\beta}R^{\mu\nu\alpha\beta}=A''^2(r)+\left(\frac{2A'(r)}{r}\right)^2
+\left(\frac{2A(r)}{r^2}\right)^2,
\end{equation}
where $A'(r)=\partial A(r)/\partial r$, and the asymptotic of metric function $A(r)$ at $r\rightarrow \infty$, for dyonic configuration, is given by Eq. (28). With the help of Eqs. (28) and (29) we find
\begin{equation}\label{30}
  \lim_{r\rightarrow\infty} K(r)=0.
\end{equation}
According to Eq. (30) spacetime becomes flat at $r\rightarrow\infty$ and there is not singularity.
It follows from the general definition (21) that even if the mass function $M(r)$ is finite at $r\rightarrow 0$ the metric function $A(r)$ is singular and according to Eq. (29) the Kretschmann scalar $K(r)$ is also singular. For the Schwarzschild BH $M(r)=m$  and  $K(0)=\infty$. For our case $M(r)$ is singular at $r=0$ and, therefore, the metric function $A(r)$ is singular. Thus, according to Eq. (29)
\begin{equation}\label{31}
 \lim_{r\rightarrow\ 0} K(r)=\infty.
\end{equation}
As a result, the Kretschmann scalar has the singularity at $r=0$. In NED models investigated in \cite{Hendi1} the singularities also hold.

\section{Thermodynamics}

The Hawking temperature is defined by
\begin{equation}
T_H=\frac{\kappa}{2\pi}=\frac{A'(r_+)}{4\pi}.
\label{32}
\end{equation}
Here, $\kappa$ is the surface gravity and $r_+$ is the event horizon. Thus, we suppose that the event horizon exists.
 Making use of Eqs. (21) and (22) we obtain the relations
\begin{equation}
A'(r)=\frac{2 GM(r)}{r^2}-\frac{2GM'(r)}{r},~~~M'(r)=r^2\rho(r),~~~M(r_+)=\frac{r_+}{2G}.
\label{33}
\end{equation}
From Eqs. (32) and (33) one finds
\begin{equation}
T_H(r_+)=\frac{1}{4\pi}\left(\frac{1}{r_+}-2Gr_+\rho(r_+)\right).
\label{34}
\end{equation}
Making use of Eqs. (24)  and (34) we obtain the Hawking temperature
\begin{equation}
T_H(z_+)=\frac{1}{4\pi\beta^{1/4}\sqrt{q_e}}\biggl(\frac{1}{z^{1/4}_+}-\frac{2Gq_e}{\sqrt{\beta}}z_+^{1/4}
f(z_+)\biggr),
\label{35}
\end{equation}
\begin{equation}
f(z)=\frac{1}{z}\sqrt{n+2z(\sqrt{z^2+1-n}-z)}-\arcsin\left(\sqrt{z^2+1-n}-z\right),
\label{36}
\end{equation}
where $z_+=r_+^4/(\beta q_e^2)$.
The event horizon $r_+$ (and $z_+$) and the parameter $\beta$ are connected by the relation $2GM(r_+)=r_+$. Then we find
\begin{equation}
\frac{2Gq_e}{\sqrt{\beta}}=\frac{4z_+^{1/4}}{\frac{4m\beta^{1/4}}{q_e^{3/2}}-\int_{z_+}^\infty\frac{f(z)}{z^{1/4}}dz}.
\label{37}
\end{equation}
Replacing unitless variable $2Gq_e/\sqrt{\beta}$ from Eq. (37) into Eq. (35) one finds the Hawking temperature of the BH in the form
\begin{equation}
T_H(z_+)=\frac{1}{4\pi\beta^{1/4}\sqrt{q_e}}\left(\frac{1}{z^{1/4}_+}-\frac{4\sqrt{z_+}f(z_+)}
{\frac{4m\beta^{1/4}}{q_e^{3/2}}-\int_{z_+}^\infty\frac{f(z)}{z^{1/4}}dz}\right).
\label{38}
\end{equation}
 Phase transitions can be investigated by studding the heat capacity. If the heat capacity is singular the second-order phase transition takes place.
It is known that the heat capacity diverges when the Hawking temperature possesses the extremum ($\partial T_H/\partial r_+=0$). Then the second-order phase transition holds.

\section{Magnetic BH}

Let us consider the magnetically charged BH.
It should be mentioned that in this case one of inequalities in Eq. (2), ${\cal L}_{{\cal F}{\cal F}}\geq 0$ is broken. Nevertheless, for completeness, we will study this case. The case of the electrically charged BH was investigated in \cite{Krug6}.
From Eq. (5) at $E=0$ we obtain the energy density
\begin{equation}
\rho(r)=\frac{1}{\beta}\arcsin\left(\frac{\beta q_m^2}{2r^4}\right).
\label{39}
\end{equation}
Then the mass function becomes
\[
M(r)=m_M-\frac{1}{\beta}\int_r^\infty r^2\arcsin\left(\frac{\beta q_m^2}{2r^4}\right)dr=m_M+\frac{r^3}{3\beta}\arcsin\left(\frac{\beta q_m^2}{2r^4}\right)
\]
\begin{equation}
-\frac{2q_m^2}{3r}F\left(\frac{1}{8},\frac{1}{2};\frac{9}{8};\frac{\beta^2q_m^4}{4r^8}\right),
\label{40}
\end{equation}
where $F\equiv {}_2F_1$ is the hypergeometric function and $m_M$ is a mass of the BH.
Making use of Eqs. (21) and (40) and introducing unitless parameter $y=(2/(\beta q_m^2))^{1/4}r$ we obtain the metric function
\[
A(y)=1-B\biggl[\frac{P}{y}+\frac{y^2}{3}\arcsin\left(\frac{1}{y^4}\right)
\]
\begin{equation}
-\frac{4}{3y^2}F\left(\frac{1}{8},\frac{1}{2};\frac{9}{8};\frac{1}{y^8}\right)\bigg],~~~B=\frac{\sqrt{2}Gq_m}{\sqrt{\beta}},
~~~P=\frac{2^{3/4}\beta^{1/4}}{q_m^{3/2}}m_M.
\label{41}
\end{equation}
Thus, we have introduced unitless parameter $P$ which characterizes the mass of the BH.
The plots of the metric function for different values of $B$ and $P$ are depicted in Figs. 1 and 2.
\begin{figure}[h]
\includegraphics[height=3.0in,width=3.0in]{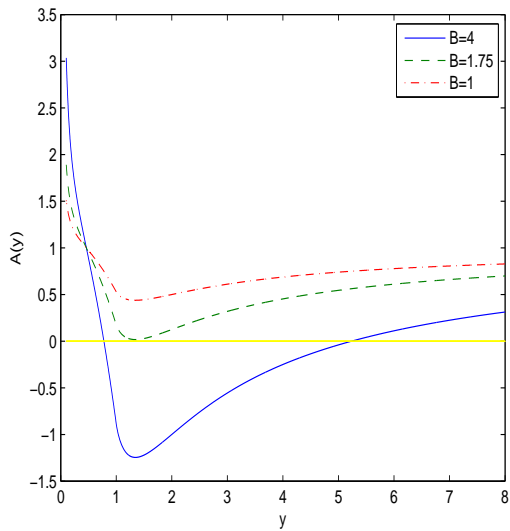}
\caption{\label{fig.1}The metric function $A(y)$ for $P=1.5$.}
\end{figure}
\begin{figure}[h]
\includegraphics[height=3.0in,width=3.0in]{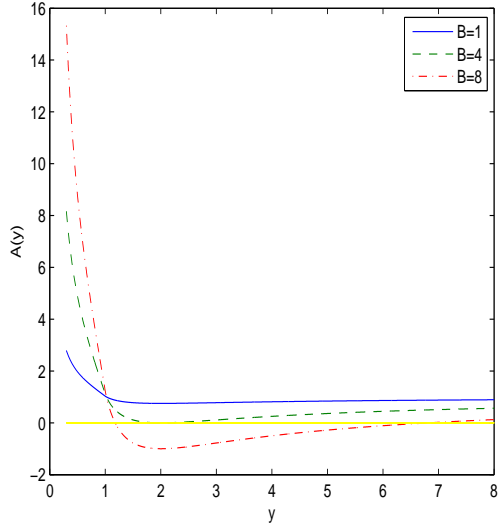}
\caption{\label{fig.2}The metric function $A(z)$ for $P=1$.}
\end{figure}
Figures 1 and 2 show that there can be BH solutions with horizons, solutions with naked singularities, and extremal BH solutions for some parameters $B$ and $P$. In accordance with Fig. 1, at $B=1$ ($P=1.5$) we have the naked singularity, at $B\approx1.75$ ($P=1.5$) one has an extremal BH solution, and at $B=4$ ($P=1.5$) there is the BH solution with two horizons. Fig. 2 shows the similar behaviour of the metric function $A(z)$ for $P=1$.

The mass function at $r\rightarrow\infty$ can be estimated making use of functions asymptotic at $r\rightarrow\infty$
\[
\frac{r^3}{3\beta}\arcsin\left(\frac{\beta q_m^2}{2r^4}\right)=\frac{q_m^2}{6r}+\frac{\beta^2 q_m^6}{144r^9}+{\cal O}(r^{-13}),
\]
\begin{equation}
\frac{2q_m^2}{3r}F\left(\frac{1}{8},\frac{1}{2};\frac{9}{8};\frac{\beta^2q_m^4}{4r^8}\right)=
\frac{2q_m^2}{3r}+\frac{\beta^2 q_m^6}{108r^9}+{\cal O}(r^{-16}).
\label{42}
\end{equation}
From Eqs. (40) and (42) we obtain the mass function at $r\rightarrow\infty$
\begin{equation}
M(r)=m_M-\frac{q_m^2}{2r}-\frac{\beta^2 q_m^6}{432r^9}+{\cal O}(r^{-13}).
\label{43}
\end{equation}
With the help of Eqs. (21) and (43) one finds the asymptotic of the metric function at $r\rightarrow\infty$,
\begin{equation}
A(r)=1-\frac{2m_MG}{r}+\frac{q_m^2G}{r^2}+\frac{\beta^2q_m^6G}{216r^{10}}+{\cal O}(r^{-14}).
\label{44}
\end{equation}
We note that corrections to the RN solution in Eq. (44) are with opposite sign comparing to the case of the electrically charged BH \cite{Krug6}.
From Eqs. (34) and (39) we obtain the Hawking temperature
\begin{equation}
T_H(r_+)=\frac{1}{4\pi}\left(\frac{1}{r_+}-\frac{2Gr_+}{\beta}\arcsin\left(\frac{\beta q_m^2}{2r_+^4}\right)\right).
\label{45}
\end{equation}
From the relation $2GM(r_+)=r_+$ and Eq. (39), one finds
\begin{equation}
\frac{2G}{\beta}=\frac{r_+}{\beta m_M-\int_{r_+}^\infty r^2\arcsin\left(\frac{\beta q_m^2}{2r^4}\right)dr}.
\label{46}
\end{equation}
Taking into consideration Eq. (46) we obtain from Eq. (45) the Hawking temperature
\begin{equation}
T_H(r_+)=\frac{1}{4\pi}\left(\frac{1}{r_+}-\frac{r_+^2\arcsin\left(\frac{\beta q_m^2}{2r_+^4}\right)}{\beta m_M-\int_{r_+}^\infty r^2\arcsin\left(\frac{\beta q_m^2}{2r^4}\right)dr}\right).
\label{47}
\end{equation}
Making use of Eq. (40) we can represent Eq. (47) as follows:
\begin{equation}
T_H(r_+)=\frac{1}{4\pi}\left(\frac{1}{r_+}-\frac{r_+^2\arcsin\left(\frac{\beta q_m^2}{2r_+^4}\right)}{\beta m_M+\frac{r_+^3}{3}\arcsin\left(\frac{\beta q_m^2}{2r_+^4}\right)
-\frac{2q_m^2\beta}{3r_+}F\left(\frac{1}{8},\frac{1}{2};\frac{9}{8};\frac{\beta^2q_m^4}{4r_+^8}\right)}\right).
\label{48}
\end{equation}
By using unitless variable $y=(2/(\beta q_m^2))^{1/4}r$ we rewrite Eq. (48) in the form
\begin{equation}
T_H(y_+)=\frac{1}{2^{7/4}\pi\beta^{1/4}\sqrt{q_m}}\left(\frac{1}{y_+}-\frac{y_+^2\arcsin\left(\frac{1}
{y_+^4}\right)}{P+\frac{y_+^3}{3}\arcsin\left(\frac{1}{y_+^4}\right)
-\frac{4}{3y_+}F\left(\frac{1}{8},\frac{1}{2};\frac{9}{8};\frac{1}{y_+^8}\right)}\right).
\label{49}
\end{equation}
Expression (49) can be used to study thermodynamics and phase transition in the magnetic BH within our model of arcsin-electrodynamics. Figures 3 and 4 depict the plots of Hawking temperature versus the variable $y_+=(2/(\beta q_m^2))^{1/4}r_+$ for different parameters $P$.
\begin{figure}[h]
\includegraphics[height=3.0in,width=3.0in]{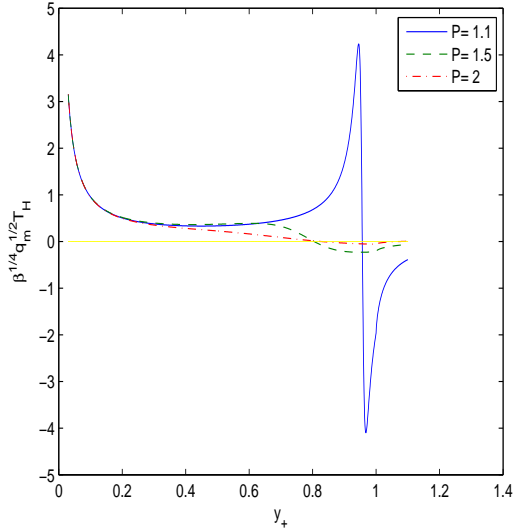}
\caption{\label{fig.3}The Hawking temperature vs. $y_+ $.}
\end{figure}
\begin{figure}[h]
\includegraphics[height=3.0in,width=3.0in]{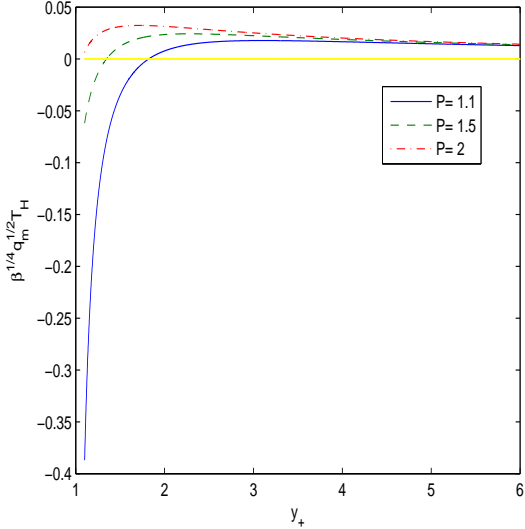}
\caption{\label{fig.4}The Hawking temperature vs. $y_+$.}
\end{figure}
Plots show that for small values of BH masses (or $P$) there are two phase transitions (one at $y_+<1$ and another at $y_+>1$) because of the extrema of the Hawking temperature. But for large masses of the BH, there is only one phase transition at $y_+>1$ and BHs are stable. For $P=1.1$ there is a singularity of the Hawking temperature at $y_+\approx 1$. If Hawking temperatures are negative, BH's are unstable.

\section{Conclusion}

We have explored arcsin-electrodynamics with free parameter $\beta$ to study dyonic and magnetic BHs. For this model the correspondence principle holds, i.e. for weak fields the model is transformed into Maxwell's electrodynamics. The corrections to Coulomb's law at $r\rightarrow\infty$ were obtained. We showed that at $q_e=q_m$ corrections disappear. Electromagnetic fields, within arcsin-electrodynamics, coupled with the gravitational field were investigated. The dyonic and magnetic solutions of the BH in GR were found.
We obtained corrections to the Reissner$-$Nordstr\"{o}m solution for $r\rightarrow\infty$, which are absent at $q_e=q_m$. The Hawking temperature of the BH was obtained. We studied the phase transitions within our model for the magnetized BH. We demonstrated that there are phase transitions for some event horizons in BHs. For the massive BH there is only one second-order phase transition.

\end{document}